\documentclass[11pt,a4paper]{article}
\pdfoutput=1

\usepackage[%
  bibstyle=numeric,%
  style=numeric-comp,%
  sorting=none,%
  sortcites=true,%
  maxnames=4,%
  abbreviate=true,%
  backend=bibtex,
]{biblatex}

\bibliography{T-duality}

\usepackage[]{Packages}

\usepackage{Definitions}

\preprint{\texttt{CERN-PH-TH/2012-224}\\\texttt{Imperial/TP/2012/LIU/01}}

\newcommand{\OfficialTitle}{Integrable Superstrings\\ on the Squashed Three-sphere}

\title{\vspace{2cm}
  {\huge   \textbf{\sffamily\OfficialTitle}}
}

\hypersetup{pdfauthor={Domenico Orlando and Linda I. Uruchurtu},pdftitle={\OfficialTitle}}

\author{
  \begin{minipage}{.8\linewidth}
    \vspace{1cm}
    \begin{center}
      {\small \textbf{Domenico Orlando}\( {}^\# \) and \textbf{Linda I. Uruchurtu}\( {}^\flat \) }
    \end{center}
    \vspace{1cm}
    \begin{minipage}{\linewidth}\centering
      {\itshape \footnotesize 
        \begin{itemize}
        \item[\( {}^\# \)] Theory Group, Physics Department, \\ Organisation européenne pour la recherche nucléaire (CERN) \\ CH-1211 Geneva 23, Switzerland
        \item[\( {}^\flat \)] The Blackett Laboratory \\ Theoretical Physics Group  \\ Imperial College London \\ 
        South Kensington, London SW7 2AZ
        \end{itemize}
      }
    \end{minipage}
  \end{minipage}
}

\date{}

\begin{document}

\begin{titlepage}

  \maketitle
  \thispagestyle{empty}

  \vfill
  \abstract{In this note we study type~\textsc{ii} superstring theory on the direct product of \( \AdS_3 \), the squashed three-sphere and a four-torus (\( \AdS_3 \times \SqS^3 \times T^4 \)). We derive explicitly the \acl{gs} action and discuss in detail the hidden integrable structure corresponding to local and non-local currents. We find that the model is classically integrable and that the currents generate a full \( psu(1,1|2) \) superalgebra, even though the corresponding spacetime isometries are broken by the squashing.}
  \vfill
\end{titlepage}
\section{Introduction}
The study of several \AdS/\CFT systems has revealed new examples of integrable systems.  Initially fuelled by
the understanding of the original gauge/gravity system, $\AdS_{5}/\CFT_{4}$, which relates string theory in
$\AdS_{5}\times S^{5}$ to $\mathcal{N}=4$ super Yang-Mills theory in four dimensions, integrability methods such as the Bethe ansatz, S--matrices, finite gap solutions, etc.~\cite{Metsaev:1998it, Bena:2003wd, Kazakov:2004qf, Beisert:2005fw, Beisert:2005bm, Beisert:2006ez, Arutyunov:2009ga}, have led to new insights into the nature of strongly coupled theories.

The same mathematical ideas can be applied to other \AdS/\CFT pairs. There has been recent progress in applications to $\AdS_{4}/\CFT_{3}$ realised as \tIIA string theory in $\AdS_{4}\times \mathbb{CP}_{3}$~\cite{Arutyunov:2008if, Stefanski:2008ik, Sorokin:2010wn} and in $\AdS_{3}/\CFT_{2}$ systems, which typically arise from the \D1/\D5 system in type~\textsc{ii} string theory~\cite{Chen:2005uj, Adam:2007ws, Babichenko:2009dk}. In both examples it has been possible to show that there exists a Lax representation of the equations of motion giving rise to an infinite set of conserved charges (local or non-local).

$\AdS_{3}/\CFT_{2}$ systems are a natural subject of study because they are severely constrained by symmetries and have target space metrics arising as near-horizon geometries of various black hole configurations~\cite{Rooman:1998xf, Detournay:2005fz, Orlando:2006cc,Compere:2008cw}.  Solutions of the form $\AdS_{3}\times \mathcal{M}$ supported by \ac{ns} fluxes have been widely studied using the technology from \textsc{wzw} models  to describe the associated worldsheet \CFT. \ac{rr}--supported geometries are more complicated, but the construction of the \ac{gs} action on $\AdS_{3}\times S^{3}$ provides evidence for their integrability. Moreover, a supercoset sigma model for backgrounds of the form $\AdS_{3}\times S^{3} \times S^{3}\times S^{1}$ and $\AdS_{3}\times S^{3}\times T^{4}$  was explicitly constructed in~\cite{Babichenko:2009dk}, and several arguments were given asserting the integrability of the model. 

Findings of novel three-dimensional target space metrics containing squashed geometries (spheres and anti-de Sitter spaces), which first appeared in the context of deformed \CFT and black holes~\cite{Rooman:1998xf, Israel:2003ry, Israel:2004vv, Israel:2004cd}, have fuelled proposals of new holographic $3D/2D$ systems based on these backgrounds~\cite{Anninos:2008fx}. Nevertheless, few studies have gone beyond obtaining the value of the central charge of the dual \CFT\footnote{A recent paper by Azeyanagi \emph{et.al}.~\cite{Azeyanagi:2012zd} studies the worldsheet theory and gives a precise formula for the spectrum of massive strings.}, also because of the absence of specific realisations in supergravity/string theory. Motivated by this, we showed in~\cite{Orlando:2010ay} how to obtain these target space metrics  from Hopf T--dualising a \D1/\D5 system sourced by \ac{rr} three-form flux with monopoles or plane waves, which were shown to preserve eight supersymmetries at the supergravity level. Metrics containing three-dimensional Schrodinger spacetime $\mathrm{Sch}_{3}$, were later shown to be obtained in a similar fashion. The construction is closely related to the realization of Melvin backgrounds~\cite{Melvin:1963qx} in string theory, pioneered by~\cite{Tseytlin:1994ei} and recently discussed in the framework of the Omega--deformation in~\cite{Hellerman:2011mv,Reffert:2011dp,Hellerman:2012zf}. It can also be understood in terms of a TsT transformation~\cite{Lunin:2005jy,Maldacena:2008wh} of the original background.

In~\cite{Ricci:2007eq}, it was argued that T--duals of integrable systems  are in turn integrable models. Notable examples are the \textsc{pcm} and $SU(2)$ sigma models~\cite{Mohammedi:2008vd,Curtright:1994be, Balog:1993es, Evans:1994hi} %
and $\AdS_{5}\times S^{5}$~\cite{Ricci:2007eq, Beisert:2008iq}, where integrability was shown to hold first for the bosonic sector, and eventually for the full superstring action. It was then natural to conjecture the existence of an integrable structure for systems involving squashed geometries, given their link to integrable models via T--duality. Reformulating the construction in~\cite{Orlando:2010ay} relating a group manifold $G$ to its squashed counterpart $\mathrm{Sq}G$\footnote{The squashing can be performed along any compact direction of a Lie Group.}, it was shown in~\cite{Orlando:2010yh} that classically, the integrable structure of the initial model is inherited by the T--dual model and  that the affine symmetry arising from the promotion of the original isometry group via the Lax construction  remains unchanged, though the zero modes are not anymore isometries of the target space\footnote{Alternatively, it is also possible to construct directly an integrable structure for the principal chiral models on squashed group manifolds as in~\cite{Kawaguchi:2010jg,Kawaguchi:2011mz,Kawaguchi:2011ub}.}.

The next step is of course to extend the previous result to the full supersymmetric model, and the Green-Schwarz (\ac{gs}) formalism is of course the natural framework to use. The \ac{gs} action can be in principle written for any supergravity background~\cite{Grisaru:1985fv,Cvetic:1999zs}. Solving the supergravity constraints order by order in the fermions, one obtains expressions for the bosonic fields that lead to the precise form of all background superfields. In practice, the procedure quickly becomes complicated, and one needs to rely on additional symmetries leading to supercoset constructions which are equivalent to the \ac{gs} action~\cite{Metsaev:1998it,Arutyunov:2008if, Stefanski:2008ik, Berkovits:1999zq}. 

In this paper we generalise the results from~\cite{Orlando:2010ay} to the full supersymmetric model. We will work with the \ac{gs} action up to quadratic order in the fermions, as the corresponding supercoset construction has not yet been realized. However, given that there is also a decoupling of flat directions (the components of the fluxes along a $T^{4}$ vanish), one can use kappa symmetry to decouple the coset-like structure by choosing an appropriate gauge. We proceed as in~\cite{Babichenko:2009dk} and discuss how to fix kappa symmetry in the T--dual model, bearing in mind that there are scenarios in which this choice is inconsistent with the dynamics, as was first discussed in~\cite{Rughoonauth:2012qd}.
We then argue that the $\AdS_{3}\times \SqS^{3}\times T^{4}$ superstring background is classically integrable by showing how to build the infinite set of conserved currents from the Killing vectors of the ten-dimensional original background and T--dualising at the level of the currents.

The plan of the paper is as follows. Section 2 is devoted to the study of the kappa-fixed \acl{gs} action of squashed backgrounds. We start by reviewing the construction of squashed backgrounds via T--duality that was proposed in~\cite{Orlando:2010ay}. We then look at the \acl{gs} superstring in a generic bosonic supergravity background up to quadratic order in fermions and write the expression for the squashed backgrounds in \tIIA theory. We end the section by discussing kappa-symmetry in the absence of a supercoset structure by looking at the flat space limit and the transformation of the vielbein under T--duality. We also give the resulting Lagrangian.
In section 3 we discuss integrability of the squashed model by looking at the currents corresponding to bosonic and fermionic symmetries. We argue that superstrings in $\AdS_{3}\times \SqS^{3} \times T^{4}$ are integrable and moreover, we show that these properties are inherited from the original (unsquashed) model. In section 4 we give a summary of our results and suggest some future directions.

\section{\acl{gs} Action of Squashed Backgrounds}
\subsection{Squashed backgrounds via T--duality}

Let us now discuss the backgrounds of interest. We will focus on \tIIB backgrounds of the form $\AdS_3 \times S^3 \times T^4$ sourced by the \ac{rr} 3-form and their Hopf T--dual backgrounds of the form $\AdS_3 \times \SqS^3 \times T^4$\footnote{Results involving warped \AdS spaces (W\AdS) and Schroedinger spaces follow analogously.}. The \tIIB solution arises as the near-horizon geometry of a system of intersecting \D1-- and \D5--branes, preserving 16 supersymmetries.  One can verify that the addition of a monopole and/or a plane wave does not alter the geometry ~\cite{Tseytlin:1996bh, Boonstra:1998yu,}. The field configuration reads
\begin{equation}
  \begin{split}
    \di s^2 ={}& Q_m Q_1^{1/2} Q_5^{1/2} \left( - \di \tau^2 + \di \omega^2
      + Q_w \di \sigma^2 + 2 \sinh \omega \di \sigma \di \tau \right) + \\
    &+ Q_m Q_1^{1/2} Q_5^{1/2} \left( \di \theta^2 + \di \phi^2 + \di
      \psi^2 + 2 \cos \theta \di \psi \di \phi \right) + \frac{Q_1^{1/2}}{Q_5^{1/2}} \left( \di y_6^2 + \dots + \di y_9^2 \right) \, , \\
    \eu^{2\Phi} ={}& \frac{Q_5}{Q_1} \, , \\
    F_{[3]} ={}& Q_m Q_1^{1/2} Q_5^{1/2}
    \left( \cosh \omega \di \tau \wedge \di \omega \wedge \di \sigma +
      \sin \theta \di \phi \wedge \di \psi \wedge \di \theta \right) \, ,
    \label{metricoriginal}
  \end{split}
\end{equation}
where $Q_{w}$ is the charge associated to the plane wave and $Q_{m}$ is the charge of the monopole. Some of the variables are periodic by construction, namely
\begin{equation}
  \begin{cases}
    \psi = \psi + 4 \pi \, ,\\
    \sigma = \sigma + 4 \pi \, ,\\
    y_i = y_i + 2 \pi \, .
  \end{cases}
\end{equation}
To obtain squashed backgrounds via T--duality, let us first introduce a new pair of $4\pi$--periodic variables $\alpha $, %
\begin{align}
  \psi = \alpha + 2 y_9 \,,
\end{align}
and rewrite the metric (\ref{metricoriginal}) as:
\begin{equation}
  \begin{split}
    \label{eq:IIB-metric}
    \di s^2 ={} & %
    \di s^2_{\AdS_{3} }
    + R^2 \left[ \di \theta^2 + \sin^2 \theta \di \phi^2 +
      \sin^2 \varpi \left(\di \alpha + \cos \theta \di \phi
      \right)^2 \right] 
    \\
    &+ \left( \di z_m + R \cos \varpi
      \left( \di \alpha + \cos \theta \di \phi \right) \right)^2   \, 
    +   4R^2 \tan^2 \varpi \left( \di y_6^2+\di y_7^2 + \di y_8^2 \right)\, ,
  \end{split}
\end{equation}
where the parameters $R$ and $\varpi$ are related to the charges by $R^2 = Q_m \sqrt{Q_1 Q_5}$  and $\cot^2 \varpi = 4 Q_m Q_5$. Also, we have rescaled $y_9$ as
\begin{align}
  z_m = 2R \cos \varpi  y_9 \, .
\end{align}
We can immediately write the T--dual
metric using Buscher's rules, interchanging $y_{9}$ for a new
coordinate $\wt y_9$, which also has periodicity $2\pi$. The metric and $B$--field become:
\begin{align}
  \wt g_{MN} &=
  {g}_{MN}+\frac{{B}_{\zeta M}{B}_{\zeta N}-{g}_{\zeta N}{g}_{\zeta M}}{{g}_{\zeta\zeta}}
  \,, & \wt {g}_{\zeta \zeta} &= \frac{(\alpha')^{2}}{{g}_{{\zeta}{\zeta}}}
  \, , & \wt g_{\zeta M} &= \alpha'
  \frac{{B}_{\zeta \sigma}}{{g}_{\zeta\zeta}} \\
  \wt B_{MN} &= {B}_{MN} +
  \frac{{B}_{\zeta M}{g}_{\zeta N}-{B}_{N\zeta}{g}_{\zeta M}}{{g}_{\zeta\zeta}}
  \, ,& \wt B_{\zeta M} &= \alpha'\frac{{g}_{\zeta M}}{{g}_{\zeta\zeta}}
  \, , & \wt \Phi &= \Phi-\frac{1}{2}\ln{\frac{{g}_{\zeta\zeta}}{\alpha'}}
  \, ,
\end{align}
where $(M,N)$ run over all coordinates except $\zeta$. However, when it comes to writing down the T--dual metric, it is convenient to introduce  a ``natural'' vielbein. Let us impose
\begin{equation}
  \wt e\indices{^a_{M}} \partial \wt X^{M} =
  e\indices{^{a}_{M}} \partial X^{M} \, ,
\label{invvielb}
\end{equation}
where $\partial X$ is the worldsheet derivative transforming under
T--duality as
\begin{align}
  \del y_9 & \to \frac{1}{g_{99}}(\alpha' \del \wt y_9 - (g_{\sigma
    9} +  B_{\sigma 9})  \del \wt X^\sigma) \,
 \qquad  \del X^\sigma \to \del \wt X^\sigma 
\end{align}
where $X^\sigma$ runs again over all the coordinates other than $u$. The
invariance of $e\indices{^a_M} \del X^M$ results in
\begin{equation}
\label{eq:vielbein-T--duality}
  \begin{dcases}
    \wt e\indices{^a_{\wt y}} = \frac{\alpha'}{g_{yy}}  e\indices{^a_{y}}\,, \\
    \wt e\indices{^a_\sigma} = e\indices{^a_\sigma} -
    \frac{g_{\sigma y } + B_{\sigma y}}{g_{yy}} e\indices{^a_{y}} & \text{for $X^\sigma \neq y_9$.}
  \end{dcases}
\end{equation}
where $y$ is a shorthand for $y_9$, which will be used from now onwards. The corresponding metric reads:
\begin{equation}
  \begin{split}
    \di s^2 ={}&  \di s_{\mathrm{\AdS}_3}^2+R^2\Big[ \di \theta^2+\sin{\theta}^2\di \phi^2+\sin^2\varpi(\di \alpha+\cos{\theta}\di \phi)^2\Big] \\
    &+\frac{\di \tilde{y}_9^2}{4R^2}\cos^2 \varpi+4R^2\tan^2
    \varpi(\di y_6^2+\di y_7^2+\di y_8^2) \, .
  \end{split}
\end{equation}
The T--dual coordinate $\wt y$ is $2 \pi$--periodic. It is convenient
to introduce the coordinate $\zeta$ dual to $z_m$ of the original metric
in Eq.~(\ref{eq:IIB-metric}) as:
\begin{equation}
  \zeta = \frac{\cos \varpi}{2R} \wt y \ ,
\end{equation}
so the metric becomes that of $\AdS_3 \times \SqS^3 \times \tilde{T}^4$. 
The \ac{rr} fields in the \tIIA theory can be obtained by reduction to nine dimensions and interpretation of the resulting expressions.  The original \tIIB \ac{rr} 3-form is given by
\begin{equation}
  F_{(3)}=R^2 ( \cosh \omega \di \tau \wedge \di \omega \wedge d\sigma + \sin \theta \di \phi \wedge \di\psi \wedge \di \theta ) \ ,
\label{fluxesIIB}
\end{equation}
whereas the \tIIA fluxes of the T--dual background are
\begin{align}
  F_{(2)}&=R \cos \varpi \sin \theta \di \theta \wedge \di \phi \ , \\
  F_{(4)}&=\left[ \omega_{\mathrm{\AdS}} +R^2 \sin^2 \varpi \sin \theta \di \theta \wedge \di \phi \wedge \di \alpha \right] \wedge \di \zeta \ , \\
  H_{(3)} &= B_2 \wedge \di \zeta = R \cos\varpi \sin \theta \di \theta \wedge \di \phi \wedge \di \zeta \ .
\label{fluxesIIA}
\end{align}

\subsection{Type II \acl{gs} superstrings in curved backgrounds}
The action for the \ac{gs} superstring in a bosonic supergravity background with constant dilaton $\Phi$, up to quadratic order in fermions is given by:
\begin{equation}
  S=-T \int \Big( \frac{1}{2}*e^Ae_A + i * e^A \bar\theta \Gamma_A \mathcal{D}\theta -i e^A \bar\theta \Gamma_A \hat\Gamma\mathcal{D}\theta \Big)+ T \int B \ ,
\label{GSaction}
\end{equation}
where $e^A(X)$, $A=0,\cdots, 9$ is the  worldsheet pullback of the vielbein, $e^A(X)={e^A}_M \partial X^M$ and 
\begin{equation}
  \hat \Gamma =
  \begin{cases}
    \Gamma_{11}  & \text{(\tIIA),} \\
    \mathbb{1}+\sigma^3 & \text{(\tIIB).} 
  \end{cases}
\end{equation}
The covariant derivative acting on the worldsheet fermions is given by
\begin{equation}
  \mathcal{D}\theta =( \nabla -\frac{1}{8} e^A \slashed{F} \Gamma_A )\theta
\end{equation}
Here $\nabla=d+\omega$ is the covariant derivative containing the spin connection of the background. The \ac{rr} fields in this expression read
\begin{equation}
  \slashed{F} =
  \begin{cases}
    -\frac{1}{2} \Gamma^{AB} \Gamma_{11}F_{AB}+\frac{1}{4!}
    \Gamma^{ABCD}F_{ABCD} & \text{(\tIIA)} \\
    i \sigma^2 \Gamma^A F_A -\frac{1}{3!} \Gamma^{ABC}F_{ABC}+\frac{i}{2\cdot 5!} \sigma^2 \Gamma^{ABCDE}F_{ABCDE} & \text{(\tIIB)}
  \end{cases}
\label{Fslashed}
\end{equation}
The worldsheet fermions can be described by two 32-component Majorana spinors in \tIIB theory of the same chirality.  In \tIIA one can consider a unique 32-component Majorana spinor, which can take the form  $\theta=\theta^{1}+\theta^{2}$ where
\begin{align}
  \theta^1 &=
  \begin{pmatrix}
    \vartheta^1 \\ 0
  \end{pmatrix} &
  \theta^2 &=
  \begin{pmatrix}
    0 \\ \vartheta^2 
  \end{pmatrix}
\end{align}
and $\vartheta^i$ are 16-component Majorana spinors of opposite chirality, $\Gamma_{11}\theta^1=\theta^1$ and $\Gamma_{11}\theta^2=-\theta^2$. After T--duality they are related to the \tIIB worldsheet spinors by~\cite{Cvetic:1999zs}
\begin{align}
  \label{eq:Tdual-theta}
  \theta_{IIB}^1 &= \theta_{IIA}^1 & \theta^{2}_{IIB}&= \gamma_y \theta_{IIA}^{2}
\end{align}
where $\gamma_y$ denotes the Dirac gamma matrix in the direction of the T--duality.

To write down the explicit form of the \ac{gs} action for the backgrounds under consideration, we need to evaluate $\slashed{F}$. Using \eqref{fluxesIIB} and \eqref{fluxesIIA}, we obtain
\begin{equation}
  \begin{cases}
    \slashed{F}_{(3)} =\frac{6}{R} (\gamma^{012}+\gamma^{345}) &\text{(\tIIB),} \\
    \slashed{F}_{(2)}=\frac{2}{R}\cos\varpi\gamma^{34} \ , \qquad   \slashed{F}_{(4)} = \frac{24}{R} (\gamma^{012y} + \gamma^{3459} \sin\varpi) 
    &\text{(\tIIA).}
  \end{cases}
\end{equation}
The Lagrangian in Eq.~\eqref{GSaction} becomes in the \tIIB case:
\begin{equation}
\mathcal{L}_{IIB} =-i\Big(\sqrt{-h}h^{ij}\delta^{IJ}-\epsilon^{ij}\sigma_3^{IJ}\Big)\bar{\theta}^I \slashed{e}_i \Big( \nabla_j \delta^{JK}+\frac{1}{48}\slashed{F}_{(3)}\slashed{e}_{j}\sigma_{1}^{JK}\Big)\theta^K
\end{equation}
with $I,J=1,2$ whereas in the (T--dual) \tIIA case:
\begin{equation}
  \begin{split}
    \mathcal{L}_{IIA} ={}& -i \bar{\theta}(\sqrt{-h}h^{ij}-\varepsilon^{ij}\Gamma_{11}){\slashed{e}_{i}}\nabla_{j}\theta+\frac{i}{8}\bar\theta(\sqrt{-h}h^{ij}-\varepsilon^{ij}\Gamma_{11})\Gamma_{11}{\slashed{e}_{i}}\gamma^{cd}
    {e_{j}}^{b}H_{bcd}\theta \\
   & -\frac{i}{16} \eu^{\Phi}\bar{\theta}(\sqrt{-h}h^{ij}-\varepsilon^{ij}\Gamma_{11})(\Gamma_{11}{\slashed{e}_{i}}\slashed{F}_{(2)}{\slashed{e}_{j}}+\frac{1}{12}{\slashed{e}_{i}}\slashed{F}_{(4)}{\slashed{e}_{j}})\theta
  \end{split}
\label{squashedGSlag}
\end{equation}
\subsection{Kappa-symmetry gauge fixing}
The \ac{gs} action can be written for any given curved background, at least to quadratic order in the fermions~\cite{Grisaru:1985fv,Cvetic:1999zs}. If one has an underlying coset structure and an underlying $\mathbb{Z}_4$ automorphism, there exists a general construction for a sigma model action for a given supergroup~\cite{Berkovits:1999zq,Babichenko:2009dk}. In the case of the supergroup $PSU(1,1\vert 2)$, the restriction to the bosonic subgroup is a \ac{gs}-type sigma model with target space $\AdS_3 \times S^3$.

At this stage it is not clear if the model can describe the motion of a superstring on a full ten-dimensional background of the form $\AdS_3 \times S^3 \times T^4$ because of  the missing torus directions that have  to be added by hand. The main issue is that in the \ac{gs} action one can find terms of the form $\bar\theta^I \partial_i X^M\Gamma_M \partial_J \theta^j$, so the torus directions necessarily couple to all the worldsheet fermions. %
However, the \ac{gs} superstring action is invariant under kappa-symmetry, local fermionic transformations of the target space coordinates. Fixing this symmetry, one can gauge away half of the fermionic degrees of freedom, so half of the \ac{gs} fermions become unphysical. Which fermions might be gauged away, of course depends on the motion of the string.

For the $\AdS_3\times S^3\times T^4$ case, it is possible to show that the $\AdS_3\times S^3$ coset action supplemented by four free bosons is equivalent to the type~\textsc{ii} \ac{gs} action in a very specific kappa symmetry gauge which sets the non-coset fermions to zero. In principle, the resulting number of fermionic degrees of freedom will be 8 (those arising from the six-dimensional coset action) as opposed to the 16 that are required in ten dimensions. However, as it was argued in~\cite{Babichenko:2009dk}, the four additional bosons interact with the coset fermions through the metric, so the kappa symmetry of the action is violated and the coset plus the $T^4$ in fact have more fermions than just the coset ones.

For backgrounds containing squashed manifolds such as $\AdS_3 \times \SqS^3 \times T^4$, the discussion of kappa-symmetry is in initially puzzling, as there is no underlying coset construction for the sigma model that might suggest which fermions to set to zero and the \ac{gs} action obtained from the Lagrangian in Eq.~\eqref{squashedGSlag} contains the same redundancy in fermionic degrees of freedom. The question then is how to fix kappa-symmetry in this case so the action correctly describes the propagation of strings. We will argue that given
that $\AdS_3\times \SqS^3 \times T^4$ can be obtained via T--duality of the $\AdS_3\times S^3 \times T^4$ model, it is possible to determine an appropriate kappa--gauge fixing condition for the \ac{gs} action. 
\subsubsection{Flat space limit}
To determine the correct kappa--gauge fixing condition for the $\AdS_3\times S^3\times T^4$ \ac{gs} action, the authors of~\cite{Babichenko:2009dk} looked at the flat space limit of the $psu(1,1\vert 2) \times psu(1,1\vert 2)$ supersymmetry algebra and identified a subalgebra of the same form within the type~\textsc{ii} flat space supersymmetry algebra. Let us review their argument. The \tIIB flat space supersymmetry algebra reads:
\begin{equation}
\left\{ q_\alpha^I, q_\beta^J \right\}=\delta^{IJ}\left[ C \gamma^M (\mathbb{1}+\Gamma_{11})\right]_{\alpha\beta}P_M
\end{equation}
with $I, J=1,2$ and $\alpha, \beta=1,\cdots 32$ are spinor indices. Introduce the projection operators
\begin{equation}
  K^{\pm}\equiv \frac{1}{2} \Big( \mathbb{1}\pm \gamma^{012345} \Big)
\label{Kprojectors}
\end{equation}
satisfying the usual properties
\begin{align}
  K^{\pm}K^{\pm} &= K^{\pm} \, , & K^{\pm} K^{\mp} &= 0 \, , & K^{\pm t} C &= C K^{\mp} \, .
\end{align}
The projected supercharges $K^{+}q^I_{\alpha}$ satisfy the flat space limit of the $psu(1,1\vert 2)$ superalgebra. Namely,
\begin{equation}
\left\{ Q^{I}_{a \alpha\dot\alpha} , Q^{J}_{b \beta\dot\beta} \right\}= \delta^{IJ} \Big[i (\varepsilon\gamma^{\mu})_{ab}\varepsilon_{\alpha\beta}\varepsilon_{\dot\alpha\dot\beta}P_{\mu}- \varepsilon_{ab}(\varepsilon\gamma^{m})_{\alpha\beta}\varepsilon_{\dot\alpha\dot\beta}P_{m}\big]
\label{psuflatsusy}
\end{equation}
where $\gamma^{\mu}=(i\sigma^{2},\sigma^{1},\sigma^{3})$ and $\gamma^{m}=(\sigma^{1},\sigma^{2},\sigma^{3})$. $P_{\mu}=S^L_\mu=S^R_\mu$, $P_{m}=L^L_n-L^R_n$, and $S_\mu$, $L_m$ are the $sl(2,\mathbb{R})$ and $su(2)$ generators, respectively. That is, there is a sub-algebra of the flat space supersymmetry algebra which has the same form as the flat space limit of $psu(2,2\vert 1)\times psu(2,2\vert 1)$. As a result, the flat space limit of the \ac{gs} action for $\AdS_3\times S^3 \times T^4$ will match the flat space \ac{gs} action in the fully fixed kappa gauge
\begin{equation}
K^{-}\theta^{I}_{IIB}=\theta^{I}_{IIB}
\label{kappagauge}
\end{equation}
\subsubsection{Kappa-gauge fixing for squashed backgrounds}
Starting from \eqref{kappagauge}, we can attempt to write down the T--dual kappa projection for the \tIIA worldsheet spinors, to determine the kappa-gauge to fix the \ac{gs} action for the $\AdS_3 \times \SqS^3 \times T^4$ background. In order for our construction to work, it is crucial to choose the vielbein described above that preserves the worldsheet derivative \( \del^a X \) after T--duality.

The gamma matrix in the direction in which the T--duality is being performed, $\gamma^{y}$, is given by
\begin{equation}
  \gamma^{y}= \cos \varpi \gamma^5 + \sin \varpi \gamma^9 \ ,
 \label{gammay}
\end{equation}
since
\begin{equation}
g_{yy}=4 R^{2}\sec^2 \varpi \ .
\end{equation}
Notice that (\ref{gammay}) depends on the deformation parameter and that
\begin{equation}
\gamma^{y}\cdot \gamma^{y}=1
\end{equation}
Using the transformation in Eq.~\eqref{eq:Tdual-theta}, the condition in \eqref{kappagauge} turns into 
\begin{align}
  \theta_{IIA}^1 &= \theta_{IIB}^1 = K^{-} \theta_{IIB}^1 = K^{-} \theta_{IIA}^1 \ , \label{kappa1} \\
  \theta_{IIA}^2 &= \gamma_y \theta_{IIB}^2 = \gamma_y K^{-} \theta_{IIB}^2 = \left( \gamma_y K^{-} \gamma_y \right) \theta_{IIA}^2 \ . \label{kappa2}
\end{align}
Since $\gamma_y $ is an involution, $\tilde{K}^+ ( \varpi ) =  \gamma_{y}K^{-}\gamma_{y}$ is the T--dual projector satisfying the usual projector properties. In detail, this projector reads
\begin{equation}
  \tilde{K}^{+}(\varpi) =\frac{1}{2}\Big( \mathbb{1}+\gamma^{01234}\Big( \cos 2\varpi \gamma^{5} + \sin 2 \varpi \gamma^{9}\Big) \Big) \equiv\frac{1}{2}\Big( \mathbb{1}+\gamma^{01234z}\Big) \ ,
  \label{kappaproj}
\end{equation}
where we introduced
\begin{equation}
  \gamma^z=\cos 2\varpi \gamma^5 + \sin 2\varpi \gamma^9 \ .
\label{gammaz}
\end{equation}
The limit $\varpi \to \pi/2$, corresponds to the case in which the deformation is zero and the geometry is $S^{3}$, so $\tilde{K}^+( \pi /2) = K^{-}$. For $\varpi\rightarrow 0$, we recover the $S^{2}\times S^{1}$ geometry, so $\tilde{K}^+(0) =  K^{+}$.

Equations \eqref{kappa1} and \eqref{kappa2} determine the kappa projections on the \tIIA worldsheet fermions such that the \ac{gs} action of $\AdS_3 \times \SqS^3 \times T^4$ in the flat space limit matches the flat space \tIIA \ac{gs} action in the given kappa-gauge. We now verify this explicitly.
The \tIIA flat space supersymmetry algebra reads
\begin{equation}
\Big\{ Q_{a}, Q_{b} \Big\}= ( C \gamma^{M} )_{ab}P_{M} \, .
\label{susyIIA}
\end{equation}
Here, $\Gamma^M$ are $32\times32$ Dirac matrices of $SO(1,9)$, $M=0\cdots 9$ is the 10-dimensional vector index and $C$ is the charge conjugation matrix. We can rewrite the algebra in terms of 16-component real Majorana spinors, $Q_\alpha \in \mathbf{16}$ and $Q_{\dot{\alpha}} \in \mathbf{16'}$.  The algebra (\ref{susyIIA}) becomes:
\begin{equation}
  \begin{split}
    \Big\{ Q_{\alpha}, Q_{\beta} \Big\} &= 2 i (\Sigma^{M} C^{-+})_{\alpha\beta}P_M \\
    \Big\{ Q_{\dot{\alpha}}, Q_{\dot{\beta}} \Big\} &= 2 i
    (\bar\Sigma^{M} C^{+-})_{\dot{\alpha}\dot{\beta}}P_M
  \end{split}
\end{equation}
where
\begin{align}
  \Gamma^{M} &=
  \begin{pmatrix}
    0 & \Sigma^{M}  \\
    \bar{\Sigma}^M & 0 
  \end{pmatrix} \, , &
  C &=
  \begin{pmatrix}
    0 & C^{+-}  \\
    C^{-+} & 0    
  \end{pmatrix} \, .
\end{align}
The $psu(1,1\vert 2)$ supersymmetry algebra reads:
\begin{equation}
  \{ Q_{a} , Q_{b}\}=i(\varepsilon \gamma^{\mu}) \otimes \varepsilon \otimes \varepsilon S_{\mu}-\varepsilon\otimes \varepsilon \gamma^{m} \otimes \varepsilon L_{m} \ ,
  \label{psususy}
\end{equation}
where the $S_{\mu}$ are the generators associated to $sl(2,\mathbb{R})$ and the $L_{m}$ are the generators associated to $su(2)$. The matrices $\gamma^{\mu}$ and $\gamma^{m}$ are defined in appendix \ref{appA:gamma}. %
After T--duality, only the $psu(1,1\vert 2)_L$ algebra is preserved. The $su(2)_{R}$ symmetry within the $psu(1,1\vert 2)_{R}$ breaks to $U(1)_{R}$. Therefore, taking the flat space limit will yield an expression similar to  \eqref{psuflatsusy}, but for the fact that now the generator associated to the preserved $u(1)$ (e.g. the T--duality circle) will appear explicitly on the right hand side. 

For the T--dual model, introduce the projector
\begin{equation}
  \Pi=K^{+}(1-\Gamma_{11})+ K^{-}(\varpi)(1+\Gamma_{11}) \ ,
\end{equation}
where $K^{+}$ and $K^{-}(\varpi)$ were defined in \eqref{Kprojectors} and \eqref{kappaproj}, respectively. To find the subalgebra of the \tIIA supersymmetry algebra that reduces, in the flat space limit, to that of broken 
$psu(1,1\vert 2)$, we evaluate:
\begin{align}
\Big\{ K^{+}Q_{\alpha}, K^{+}Q_{\beta} \Big\} &=2i (\Sigma^M C^{-+})_{\dot{\alpha}\dot{\beta}}P_M \ ,&
\Big\{ K^{-}(\varpi)Q_{\dot{\alpha}}, K^{-}(\varpi)Q_{\dot{\beta}} \Big\} &= 2i (\Sigma^M C^{+-})_{\dot{\alpha}\dot{\beta}}P_M \ .
\label{calcsusy}
\end{align}
Using the basis in appendix \ref{appA:title}, we find that the first anticommurator in \eqref{calcsusy} is
\begin{equation}
\Big\{ K^{+}Q, K^{+}Q \Big\}= (1+\sigma^3) \otimes \Big( i\varepsilon \gamma^\mu \otimes \varepsilon \otimes \varepsilon P_{\mu} + \varepsilon \otimes \varepsilon \gamma^{m}\otimes \varepsilon P_m \Big)
\end{equation}
where $\mu=0, 1, 2$ and $m=3, 4, 5$. The subset of projected supercharges $K^{+}Q$ has the same anticommutation relation as the flat space limit of $psu(1,1\vert 2)$, so for the part of the action involving the $\theta^{1}$ spinors,  will match the flat space \ac{gs} action in the kappa gauge
\begin{equation}
K^{-}\theta^{1}=\theta^{1}
\end{equation}
found in Eq.\eqref{kappa1}. The generators associated to the $T^4$ do not appear in the right hand side of the anticommutator, and the torus decouples.

The second anticommutator in \eqref{calcsusy} gives:
\begin{align}
\Big\{ \tilde{K}^{-}(\varpi)Q,\tilde{K}^{-}(\varpi)Q \Big\}&={}(\mathbb{1}-\cos 2\varpi \sigma_{3})\otimes \Big[ i\varepsilon \gamma^{\mu}\otimes \varepsilon \otimes \varepsilon P_{\mu} \Big] - \sin 2\varpi \Big[ \varepsilon \otimes \varepsilon \gamma^{\mu}\otimes \sigma_{1}\otimes \varepsilon  P_{\mu}\Big] \nonumber
\\
&+(\sigma^{3}-\cos 2\varpi \mathbb{1})\otimes \Big[ \varepsilon \otimes \varepsilon \gamma^{m} \otimes \varepsilon P_{m}\Big]+\sin 2\varpi  \Big[ i \sigma^{1} \otimes \varepsilon \otimes \sigma^{1}\gamma^{m}\otimes \varepsilon P_{m}\Big]
\nonumber \\
&+(\mathbb{1}-\cos 2\varpi \sigma^{3}) \otimes \Big[ \varepsilon \otimes \sigma^{1} \otimes\varepsilon P_{z}\Big] + \sin 2\varpi \Big[ -i\varepsilon \otimes \varepsilon \otimes \varepsilon \otimes \varepsilon P_{z}\Big] \nonumber \\
\end{align}
Here $m=3,4$. The presence of \( P_z \) in the right hand side signals the breaking of the $psu(1,1\vert 2)_{R}$ symmetry. Following the same logic as before, the subset of projected supercharges $\tilde{K}^{-}(\varpi)Q$ are related to the $\theta^{2}$ worldsheet spinors,  so the part of the action involving the $\theta^{2}$ spinors will match the flat space \ac{gs} action in the kappa gauge
\begin{equation}
  \tilde{K}^{+}(\varpi)\theta^{2} = \theta^{2} \ ,
\end{equation}
which is just the condition derived using T--duality in \eqref{kappa2}. Notice that also in this case, the directions $6,7,8$ and $\bar{z}$ (orthogonal to the $z$ direction) specify a $\tilde{T}^{4}$ which decouples from the $\AdS_{3}\times \SqS^{3}$.
\subsection{The $\AdS_{3}\times \SqS^{3} \times T^{4}$ case }
Having determined the kappa-gauge fixing condition required to obtain a \ac{gs} action which i). reduces to the familiar \tIIA \ac{gs} action in the flat space limit; ii). projects out the orthogonal $T^{4}$ to the $\AdS_{3}\times \SqS^{3}$ space; and iii). has the correct number of fermionic degrees of freedom, we are in the position to write down the explicit expression for the \ac{gs} action. Let us start from \eqref{squashedGSlag}, and focus on the \ac{ns} sector. 

\subsubsection{The \acl{ns} sector}
Consider the squashed sphere subsector with the additional $S^1$. We can read off the relevant vielbein components from \eqref{tdualvielbein}:
\begin{equation}
  \begin{aligned}
    e^{3}&=R \di \theta \ , & e^4=R\sin(\theta) \di \varphi \ ,\\
    e^5&=\frac{\cos^2{\varpi}\di y}{2R}+R\sin^2{\varpi}(\di \alpha+\cos{\theta}\di \varphi) \ ,  \\
    e^9&=\frac{\cos{\varpi}\sin{\varpi} \di
      y}{2R}-R\sin{\varpi}\cos{\varpi}(\di \alpha+\cos{\theta}\di
    \varphi) \ ,
  \end{aligned}
\end{equation}
It is convenient to use the rotated vielbein $\{e^3,e^4,e^5,e^9\}\rightarrow \{e^3,e^4,\tilde{e}^5,\tilde{e}^9\}$ where
\begin{equation}
  \tilde{e}^{5}=-R\sin{\varpi}(\di \alpha+\cos{\theta}\di \varphi) \qquad \tilde{e}^9=\frac{\cos{\varpi}}{2R}\di y \ .
\end{equation}
The three-form flux reads
\begin{equation}
  H_3 = \frac{\cos{\varpi}}{R} \tilde{e}^{9}\wedge e^3 \wedge e^4 \ ,
\label{s3form}
\end{equation}
and the relevant spin connection components are:
\begin{equation}
  \begin{aligned}
    \omega^{4\tilde{5}}&=\frac{1}{2} \sin{\varpi} d\theta \ ,& \omega^{3\tilde{5}}=-\frac{1}{2}\sin{\varpi}\sin{\theta}d\varphi \ , \\
    \omega^{34}&=\frac{1}{2}\sin^2{\varpi}d\alpha-\frac{1}{2}\cos{\theta}(1+\cos^2{\varpi})d\varphi \ .
  \end{aligned}
\label{sqsc}
\end{equation}
In what follows, we will be using the following expressions:
\begin{equation}
  \begin{aligned}
    \gamma^{\tilde{5}}&\equiv \gamma^{\bar{y}}=-\sin{\varpi}\gamma^5+\cos{\varpi}\gamma^9 \ , \\
    \gamma^{\tilde{9}}&\equiv \gamma^{y}=\cos{\varpi}\gamma^5+\sin{\varpi}\gamma^9 \ , \\
    \gamma^{z}&=\cos{2\varpi}\gamma^5+\sin{2\varpi}\gamma^9 = \cos{\varpi}\gamma^y+\sin{\varpi}\gamma^{\bar{y}} \ , \\
    \gamma^{\bar{z}}&=-\sin{2\varpi}\gamma^5+\cos{2\varpi}\gamma^9 =
    -\sin{\varpi}\gamma^y+\cos{\varpi}\gamma^{\bar{y}} \ .
  \end{aligned}
\end{equation}
Introducing the kappa-projectors, we can write down the \ac{gs} Lagrangian for the \ac{ns} sector
\begin{multline}
\mathcal{L}_{NS}=\left\{-i\bar{\theta}^1 (\sqrt{h}h^{ij}-\varepsilon^{ij}\Gamma_{11})K^{+}+i\bar{\theta}^2(\sqrt{h}h^{ij}-\varepsilon^{ij}\Gamma_{11})K^{-}(\varpi)\right\}\times
 \\
 \left[ \slashed{e}_i D_j -\frac{1}{8}\Gamma_{11}\slashed{e}_i \gamma^{cd}e_j^b H_{bcd} \right]\left\{ K^{-}\theta^1 + K^{+}(\varpi)\theta^2\right\} \ .
 \label{LNSs}
\end{multline}
Using (\ref{s3form}) and (\ref{sqsc}), the term in square brackets in (\ref{LNSs}) becomes:
\begin{equation}
  \begin{aligned}
    &\slashed{e}_i \left[ \partial_j + \frac{1}{4R}\gamma^{4z}{e_j}^3- \frac{1}{4R}\gamma^{3z}{e_j}^4+ \frac{1}{4R}\gamma^{34}{e_j}^5- \frac{1}{2R}\cot{\theta}\gamma^{34}{e_j}^4 \right]K^{-}\theta^1  \\
    &\slashed{e}_i \left[ \partial_j - \frac{1}{4R}\gamma^{45}{e_j}^3+
      \frac{1}{4R}\gamma^{35}{e_j}^4- \frac{1}{4R}\gamma^{34}{e_j}^z-
      \frac{1}{2R}\cot{\theta}\gamma^{34}{e_j}^4
    \right]K^{+}(\varpi)\theta^2 \, ,
  \end{aligned}
\end{equation}
and (\ref{LNSs}) can be rewritten as

\begin{small}  
\begin{equation}
  \begin{split}
    4R\mathcal{L}_{NS}={}&-i\bar{\theta}^1 (\sqrt{h}h^{ij}-\varepsilon^{ij}\Gamma_{11})K^{+}\slashed{e}_i \left[ 4R\partial_j + \gamma^{4z}{e_j}^3- \gamma^{3z}{e_j}^4+ \gamma^{34}\Big({e_j}^5- 2\cot{\theta}{e_j}^4\Big) \right]K^{-}\theta^1  \\
    & +i\bar{\theta}^2(\sqrt{h}h^{ij}-\varepsilon^{ij}\Gamma_{11})K^{-}(\varpi)\slashed{e}_i \left[ 4R \partial_j - \gamma^{45}{e_j}^3+ \gamma^{35}{e_j}^4- \gamma^{34}\Big({e_j}^z- 2\cot{\theta}{e_j}^4\Big) \right]K^{+}(\varpi)\theta^2  \\
    &-i\bar{\theta}^1 (\sqrt{h}h^{ij}-\varepsilon^{ij}\Gamma_{11})K^{+}\slashed{e}_i \left[ 4R \partial_j - \gamma^{45}{e_j}^3+ \gamma^{35}{e_j}^4- \gamma^{34}\Big({e_j}^z- 2\cot{\theta}{e_j}^4\Big) \right]K^{+}(\varpi)\theta^2  \\
    &+i\bar{\theta}^2(\sqrt{h}h^{ij}-\varepsilon^{ij}\Gamma_{11})K^{-}(\varpi)\slashed{e}_i
    \left[ 4R \partial_j + \gamma^{4z}{e_j}^3- \gamma^{3z}{e_j}^4+
      \gamma^{34}\Big({e_j}^5- 2\cot{\theta}{e_j}^4\Big)
    \right]K^{-}\theta^1 \, .
  \end{split}
\end{equation}
\end{small}
We need to move around the $K$-projectors and simplify the resulting expression. After some rather tedious algebra we obtain the final expression for the \ac{gs} Lagrangian on the squashed three-sphere:
\begin{small}
  \begin{equation}
    \begin{split}
      \mathcal{L}^{\SqS^{3}}_{NS} ={} &-i\bar{\theta}^1 (\sqrt{h}h^{ij}-\varepsilon^{ij}\Gamma_{11}) \Big\{ \sum^{5}_{a = 0}{e_i}^a \gamma_a  \left[ \partial_j  -\frac{1}{2R}\gamma^{5[4}{e_{j}}^{3]}\cos{2\varpi}+\frac{1}{4R}\gamma^{34}\Big({e_j}^5-2\cot{\theta}{e_j}^4 \Big)\right]  \\
      & \sum^{9}_{m = 6}{e_i}^m \gamma_m \left[ \partial_j -\frac{1}{2R}\gamma^{9[4}{e_{j}}^{3]}\sin{2\varpi}\right]\Big\}\theta^1  \\
      &+i\bar{\theta}^2(\sqrt{h}h^{ij}-\varepsilon^{ij}\Gamma_{11})\Big\{
      \sum^{z}_{a = 0}{e_i}^a \gamma_a \left[ \partial_j +
        \frac{1}{2R}\gamma^{z[4}{e_j}^{3]}\cos{2\varpi}-\frac{1}{4R}\gamma^{34}\Big({e_j}^{z}
        - 2\cot{\theta}{e_j}^{4}\Big)\right]
      \\
      &+ \sum^{\bar{z}}_{m = 6}{e_i}^m \gamma_m \left[ \partial_j -\frac{1}{2R}\gamma^{\bar{z}[4}{e_{j}}^{3]}\sin{2\varpi}\right]\Big\}\theta^2  \\
      &-i\bar{\theta}^1(\sqrt{h}h^{ij}-\varepsilon^{ij}\Gamma_{11})\Big\{ \sum^{5}_{a = 0}{e_i}^a \gamma_a  \left[ \partial_j +\frac{1}{4R}(\gamma^{5[4}{e_j}^{3]}-\gamma^{z[4}{e_j}^{3]})-\frac{1}{4R}\gamma^{34}\Big(\frac{1-\gamma^{5z}}{2}\Big)\Big( {e_j}^{z}-2\cot{\theta}{e_j}^4\Big) \right]  \\
      &+\sum^{9}_{m = 6}{e_i}^m \gamma_a \left[ \partial_j
        +\frac{1}{4R}(\gamma^{5[4}{e_j}^{3]}+\gamma^{z[4}{e_j}^{3]})-\frac{1}{4R}\gamma^{34}\Big(\frac{1+\gamma^{9\bar{z}}}{2}\Big)\Big(
        {e_j}^{z}-2\cot{\theta}{e_j}^4\Big) \right] \Big\}
      \theta^2  \\
      &+i\bar{\theta}^2(\sqrt{h}h^{ij}-\varepsilon^{ij}\Gamma_{11})\Big\{
      \sum^{z}_{a = 0}{e_i}^a \gamma_a \left[ \partial_j +
        \frac{1}{4R}(\gamma^{5[4}{e_j}^{3]}-\gamma^{z[4}{e_j}^{3]})
        +\frac{1}{4R}\gamma^{34}\Big(\frac{1+\gamma^{9\bar{z}}}{2}\Big)\Big( {e_j}^5-2\cot{\theta}{e_j}^4 \Big )\right]  \\
      &+ \sum^{\bar{z}}_{m = 6}{e_i}^a \gamma_a \left[ \partial_j -
        \frac{1}{4R}(\gamma^{5[4}{e_j}^{3]}+\gamma^{z[4}{e_j}^{3]})
        +\frac{1}{4R}\gamma^{34}\Big(\frac{1-\gamma^{5z}}{2}\Big)\Big(
        {e_j}^5-2\cot{\theta}{e_j}^4 \Big )\right]\Big\}\theta^1
    \end{split}
  \end{equation}
\end{small}
The \AdS sector is simpler. Using \eqref{spinconn} it is easy to evaluate the contribution coming from the spin connection:
\begin{align}
\slashed{e}_{\mu}&= \frac{1}{4}\omega_j^{AB}\gamma_{AB} \nonumber \\
&=\frac{1}{2}\left( -\frac{{e_j}^{2}}{2R}\gamma^{01}+\frac{{e_j}^{0}}{R}( \tanh {\omega}\gamma^{01}-\frac{1}{2}\gamma^{12})-\frac{{e_j}^1}{2R}\gamma^{02}\right) \ ,
\end{align}
and expanding the kappa-projectors, we write down the Lagragian as:
\begin{small}
  \begin{equation}
    \begin{aligned}
      \mathcal{L}^{\mathrm{\AdS}_{3}}_{NS}&=-i\bar{\theta}^1
      (\sqrt{h}h^{ij}-\varepsilon^{ij}\Gamma_{11}) \Big\{ \sum^{5}_{a
        = 0}{e_i}^a \gamma_a \left[ \partial_j - \frac{1}{4}({e_j}^2
        -2 \tanh
        {\omega}{e_j}^0)\gamma^{01}-\frac{{e_j}^0}{4R}\gamma^{12}-\frac{{e_j}^1}{4R}\gamma^{02}\right]\Big\}\theta^1
       \\
      &+\bar{\theta}^2 (\sqrt{h}h^{ij}-\varepsilon^{ij}\Gamma_{11})
      \Big\{ \sum^{\bar{z}}_{m = 6}{e_i}^m \gamma_m \left[ \partial_j
        - \frac{1}{4}({e_j}^2 -2 \tanh
        {\omega}{e_j}^0)\gamma^{01}-\frac{{e_j}^0}{4R}\gamma^{12}-\frac{{e_j}^1}{4R}\gamma^{02}\right]\Big\}\theta^2
       \\
      &+\bar{\theta}^2 (\sqrt{h}h^{ij}-\varepsilon^{ij}\Gamma_{11}) \Big\{ \sum^{z}_{a = 0}{e_i}^a \gamma_a  \left[ \partial_j - \frac{1}{4}({e_j}^2 -2 \tanh {\omega}{e_j}^0)\gamma^{01}-\frac{{e_j}^0}{4R}\gamma^{12}-\frac{{e_j}^1}{4R}\gamma^{02}\right]\Big(\frac{1+\gamma^{9\bar{z}}}{2}\Big)  \\
      &+ \sum^{\bar{z}}_{m = 6}{e_i}^m \gamma_m \left[ \partial_j -
        \frac{1}{4}({e_j}^2 -2 \tanh
        {\omega}{e_j}^0)\gamma^{01}-\frac{{e_j}^0}{4R}\gamma^{12}-\frac{{e_j}^1}{4R}\gamma^{02}\right]\Big(\frac{1+\gamma^{5z}}{2}\Big)\Big\}
      \theta^1 
       \\
      &-i\bar{\theta}^1 (\sqrt{h}h^{ij}-\varepsilon^{ij}\Gamma_{11}) \Big\{ \sum^{5}_{a = 0}{e_i}^a \gamma_a  \left[ \partial_j - \frac{1}{4}({e_j}^2 -2 \tanh {\omega}{e_j}^0)\gamma^{01}-\frac{{e_j}^0}{4R}\gamma^{12}-\frac{{e_j}^1}{4R}\gamma^{02}\right]\Big(\frac{1-\gamma^{5z}}{2}\Big)  \\
      &+ \sum^{9}_{m = 6}{e_i}^m \gamma_m \left[ \partial_j -
        \frac{1}{4}({e_j}^2 -2 \tanh
      {\omega}{e_j}^0)\gamma^{01}-\frac{{e_j}^0}{4R}\gamma^{12}-\frac{{e_j}^1}{4R}\gamma^{02}\right]\Big(\frac{1+\gamma^{9\bar{z}}}{2}\Big)\Big\}
      \theta^2
    \end{aligned}
  \end{equation}
\end{small}
\subsubsection{The \acl{rr} sector}
We now turn to the \ac{rr} sector. We are interested in expressing the combination
\begin{equation}
\Gamma_{11} \slashed{e}_i \slashed{F}_2 \slashed{e}_j + \frac{1}{12}\slashed{e}_i \slashed{F}_4 \slashed{e}_j
\end{equation}
in terms of the kappa-symmetry projectors $K^{\pm}$ and $K^{\pm}(\varpi)$. The relevant part of the Lagrangian is:
\begin{multline} 
  \mathcal{L}_{\ac{rr}} = \Big\{ -i \bar{\theta}^1 ( \sqrt{h}h^{ij} - \varepsilon^{ij} \Gamma_{11} ) K^{+} + i \bar{\theta}^2 ( \sqrt{h} h^{ij} - \varepsilon^{ij} \Gamma_{11} ) K^{-}(\varpi) \Big \} \times  \\
\frac{1}{16} \Big( \Gamma_{11} \slashed{e}_i \slashed{F}_2 \slashed{e}_j + \frac{1}{12}\slashed{e}_i \slashed{F}_4 \slashed{e}_j \Big)\Big\{K^{-}\theta^1+K^{+}(\varpi)\theta^2\Big\} \ .
\label{LRRs}
\end{multline}
Consider the $\bar{\theta}^1  \theta^1$-term. Since $\bar{\theta}^1 \Gamma_{11} = - \bar{\theta}^1$, we can rewrite (\ref{LRRs}) as
\begin{equation}
-i\bar{\theta}^1(\sqrt{h}h^{ij}-\varepsilon^{ij}\Gamma_{11})K^{+}
\Big[ \frac{1}{4R}\slashed{e}_i \gamma^{012y} K^{+} \slashed{e}_j  \Big]K^{-}\theta^1 \ .
\end{equation}
Expanding the projectors, the previous expression becomes:
\begin{equation}
- \frac{i}{4R}\bar{\theta}^1(\sqrt{h}h^{ij}-\varepsilon^{ij}\Gamma_{11})
\Big[\sum^5_{a=0} e^a_i \gamma_a \sin{\varpi}\gamma^{0129}  + 
\sum^9_{m=6} e^m_i \gamma_m \cos{\varpi}\gamma^{0125}
\Big]\sum^5_{b=0}e^b_j \gamma_b \theta^1 \ .
\end{equation}
We proceed analogously with the $\bar{\theta}_2 \theta_2$-term. Rewriting the fluxes
\begin{equation}
i\bar{\theta}^2(\sqrt{h}h^{ij}-\varepsilon^{ij}\Gamma_{11})K^{-}(\varpi)
\Big[ \frac{1}{4R}\slashed{e}_i \gamma^{012y} K^{-}(\varpi) \slashed{e}_j  \Big]K^{+}(\varpi)\theta^2 \ ,
\end{equation}
and acting with the projectors leaves
\begin{equation}
 \frac{i}{4R}\bar{\theta}^2(\sqrt{h}h^{ij}-\varepsilon^{ij}\Gamma_{11})
\Big[-\sum^{z}_{a=0} e^a_i \gamma_a \sin{\varpi}\gamma^{012\bar{z}}  + \sum^{\bar{z}}_{m=6} e^m_i \gamma_m \cos{\varpi}
\gamma^{012z}
\Big]\sum^5_{b=z}e^b_j \gamma_b \theta^2 \ .\\
\end{equation}
Finally, the crossed terms use the same expressions we obtained for the fluxes in terms of the kappa-projectors. The final results read:
\begin{small}
\begin{multline}
- \frac{i}{4R}\bar{\theta}^1(\sqrt{h}h^{ij}-\varepsilon^{ij}\Gamma_{11})\Big\{
\sum^5_{a=0} e^a_i \gamma_a \Big[ \frac{1}{2}\sin{\varpi}\gamma^{0129}\sum^{\bar{z}}_{b=0}e^b_j \gamma_b  +
 \frac{1}{2}\sin{\varpi}\gamma^{012\bar{z}}(\sum^{\bar{z}}_{a=6}e^a_j \gamma_a-\sum^{z}_{a=0}e^a_j \gamma_a)\Big] \\
+\sum^9_{m=6} e^m_i \gamma_m \Big[ \frac{1}{2}\cos{\varpi}\gamma^{0129}\sum^{\bar{z}}_{b=0}e^b_j \gamma_b 
+\frac{1}{2}\cos{\varpi}\gamma^{012z}(\sum^{z}_{a=0}e^a_j \gamma_a-\sum^{\bar{z}}_{a=6}e^a_j \gamma_a)\Big]\Big\} \theta^2 \ ,
\end{multline}
\end{small}
and
\begin{small}
\begin{multline}
\frac{i}{4R}\bar{\theta}^2(\sqrt{h}h^{ij}-\varepsilon^{ij}\Gamma_{11})\Big\{
\sum^z_{a=0} e^a_i \gamma_a \Big[- \frac{1}{2}\sin{\varpi}\gamma^{012\bar{z}}\sum^{\bar{z}}_{b=0}e^b_j \gamma_b  +
 \frac{1}{2}\sin{\varpi}\gamma^{0129}(\sum^{5}_{a=0}e^a_j \gamma_a-\sum^{9}_{a=6}e^a_j \gamma_a)\Big]  \\
+\sum^{\bar{z}}_{m=6} e^m_i \gamma_m \Big[ \frac{1}{2}\cos{\varpi}\gamma^{012z}\sum^{\bar{z}}_{b=0}e^b_j \gamma_b 
+\frac{1}{2}\cos{\varpi}\gamma^{0125}(-\sum^{5}_{a=0}e^a_j \gamma_a+\sum^{9}_{a=6}e^a_j \gamma_a)\Big]\Big\} \theta^1 \ .
\end{multline}
\end{small}

We do not have anymore a supercoset structure for the squashed background. However, as T--duality makes  apparent, there is a natural separation in the fermionic coordinates in those coming from torus coordinates $(6, 7, 8,\bar{z})$ and the remaining ones. Following the notation in~\cite{Rughoonauth:2012qd}, we write:
\begin{equation}
  \begin{aligned}
    \zeta^{1}&=K^{-}\theta^{1} \, , & v^{1} &= K^{+}\theta^{1} \ , \\
    \zeta^{2}&=\tilde{K}^{+}(\varpi)\theta^{2} \, , &     v^{2} &= \tilde{K}^{-}(\varpi)\theta^{2} \ .
  \end{aligned}
\end{equation}
The vielbein splits as
\begin{equation}
\slashed{e}_{i}={\tilde{e}_{i}}^{a}\gamma_{a}+\partial_{i}y^{m}\gamma_{m} \ ,
\end{equation}
where $m$ runs over the torus coordinates. The torus fields $y$ never appear linearly in the Lagrangian, and can be set to zero by admitting the trivial solution to the equations of motion,  $y^{i,m}=v^{i}=0$. Hence, we can reduce the theory to the $\AdS_{3}\times \SqS^{3}$ sector coupled to sixteen fermions, eight from $\zeta^{1}$ and eight from $\zeta^{2}$. Using kappa-symmetry, the resulting number of physical degrees of freedom is eight, and the resulting \ac{gs} Lagrangian is given by
\begin{equation}
  \begin{split}
    \mathcal{L}_{GS} ={}&
    -\bar\zeta^{1}(\sqrt{-h}h^{ij}-\varepsilon^{ij}\Gamma_{11})\slashed{\tilde{e}}_{i}\left[
      D_{j}+\frac{1}{8}\gamma^{cd}e_{j}^{b}H_{bcd}+\frac{1}{4R}\gamma^{012y}K^{+}\slashed{\tilde{e}}_{j}\right]\zeta^{1}
    \\
    &-\bar\zeta^{2}(\sqrt{-h}h^{ij}-\varepsilon^{ij}\Gamma_{11})\slashed{\tilde{e}}_{i}\left[ D_{j}+\frac{1}{8}\gamma^{cd}e_{j}^{b}H_{bcd}+\frac{1}{4R}\gamma^{012y}\tilde{K}^{-}(\varpi)\slashed{\tilde{e}}_{j}\right]\zeta^{2}
  \end{split}
\end{equation}
and by construction, this theory will be T--dual to the $psu(1,1\vert 2)\times psu(1,1\vert 2) / su(1,1)\times su(2)$ sigma model. As before, kappa-symmetry gauges away half of the components of the worldsheet fermions. 

Our specific gauge choice can only describe some string configurations. For example, the corresponding gauge in the undeformed case cannot describe a motion restricted to the $\AdS_3 \times S^3 $ subspace~\cite{Rughoonauth:2012qd}. We expect a similar type of limitation for our model.

As we have seen, despite the lack of supercoset structure in the squashed case, it is possible to construct the kappa-fixed GS-action. Interestingly enough, the T--dual result inherits most of the characteristics of the original model, with the orthogonal torus being projected out. In the next section, we will argue that in fact, integrability is also inherited from the original model and that T--duality preserves the structure.
\section{Currents and integrability}
\label{sec:curr-integr}
Having constructed the \ac{gs} Lagrangian for our model, we can derive the Noether currents corresponding to the bosonic and fermionic symmetries using a standard procedure. This would be straightforward but incomplete. As already emphasized in~\cite{Orlando:2010yh}, the sigma model with squashed sphere target space inherits \emph{all the symmetries} of the T--dual model on the round sphere \( S^3 \) even though the background metric only has part of the isometries of the round sphere. The extra symmetries correspond to non-local currents that together with the Noether ones realise the full \( psu(1,1|2) \)  superalgebra, even if they \emph{do not} stem from local symmetries.
We will employ the following strategy. First we calculate the currents for the T--dual system with \( S^3 \) target space and then we transform them using T--duality.

Consider the \ac{gs} action for \( \AdS_3 \times S^3 \times T^4 \), as written in~\cite{Babichenko:2009dk}. Following~\cite{Sorokin:2011rr} it is possible to write the explicit expression for the Noether currents as
\begin{equation}
  J = J^A K_A + J^{AB}  \nabla_A K_B \ ,
\end{equation}
where \( K_A \) are the Killing vectors of the ten-dimensional background
\begin{align}
  J^A &= e^A + i \theta \Gamma^A \mathcal{D} \theta - \frac{i}{8} \Gamma^A \slashed{F} \Gamma_B \theta + i \theta \Gamma^A \hat \Gamma * \mathcal{D} \theta - \frac{i}{8} * e^B \theta \Gamma^A \hat \Gamma \slashed F \Gamma_B \theta \ , \\
  J^{AB} &= - \frac{i}{4} e^C \theta {\Gamma^{AB}}_C \theta + \frac{i}{4} * e^C \theta {\Gamma^{AB}}_C \hat \Gamma \theta \ ,
\end{align}
These currents are conserved on-shell and generate the \( psu(1,1|2) \) algebra. Using a standard procedure this is the starting point for the construction of an infinite tower of conserved charges which assure the classical integrability of the model based on the preserved \( \setZ^4 \) grading~\cite{Bena:2003wd}.

Let us now go back to our squashed-sphere model. As already discussed, T--duality breaks half of the local symmetries. On the bosonic side this is reflected in the fact that currents that do not commute with the generator \( J_3 \) of the \( su(2) \oplus su(2) \) symmetry are not anymore local currents, but turn into non-local currents as explained in~\cite{Orlando:2010yh}. We distinguish two cases: 
\begin{enumerate}
\item For the currents that commute with \( J_3 \), bosonic T--duality is implemented via the substitution 
  \begin{equation}
    \di z \mapsto * \di \zeta - R \cos \varpi \left( \di \alpha + \cos \theta \di \phi \right)  \, .
  \end{equation}
  In other words, the expression for the currents remain formally the same as in the equation above, but the vielbein \( e^A \) are transformed as follows: 
  \begin{equation}
    \begin{cases}
      {e_m}^A \di x^\mu  \mapsto {e_\mu}^A \di x^\mu & \text{for \(\mu \neq z\),} \\
      {e_z}^A \di z  \mapsto {e_z}^A \left( * \di \zeta - R \cos \varpi \left( \di \alpha + \cos \theta \di \phi \right) \right) \, .
    \end{cases}
  \end{equation}
\item The currents that \emph{do not commute} with \( J_3 \) have an explicit dependence on the coordinate \( z \) (as opposed to just depending on the differential \( \di z  \)) and require an extra step. Concretely we need to change the gauge as follows
  \begin{equation}
    J' = h^{-1} J h + h^{-1} \di h \ ,
  \end{equation}
  where
  \begin{equation}
    h = \exp [ -i \left( \alpha + \cos \varpi \, z \right) T_3] \ ,
  \end{equation}
and then apply the transformation on the vielbein. The resulting currents are non-local but are conserved and satisfy the usual commutation relations.%
\end{enumerate}

In the previous section we have shown that it is possible to fix kappa-symmetry such that the \( T^4 \) is decoupled from the \( \AdS_3 \times \SqS^3 \) part. This means that in this gauge we are back to the supercoset description of~\cite{Bena:2003wd}, which admits a \( \setZ^4 \) structure preserved by T--duality. The currents obtained above are conserved by construction and even if non-local, they can be used to obtain infinite towers of conserved charges. It follows that our model is still classically integrable.

\section{Summary and Future Directions}
In this paper we have considered \tIIA \ac{gs}  superstrings in  $\AdS_{3}\times \SqS^{3} \times T^{4}$ backgrounds which are obtained via Hopf T--duality of a \D1/\D5/monopole background along a hybrid direction mixing a torus coordinate with an $S^{3}$ coordinate.  We have explicitly constructed the Lagrangian up to quadratic order in fermions and have discussed the fixing of kappa-symmetry and shown that it is possible to determine an appropriate kappa--gauge fixing condition that decouples the $T^{4}$ components from the $\AdS_{3}\times \SqS^{3}$ sector.

Despite the fact that our squashed background does not possess a supercoset structure after T--duality, a careful choice of the dual vielbein allows the inheritance of the original model's properties, with kappa-gauge fixing removing half of the components of the worldsheet fermions. 

The background is also classically integrable, with the properties being inherited from the original model. For this, we have shown how to build an infinite set of conserved currents, which realise the full $psu(1\vert 1,2)$ superalgebra, where the non-local currents are associated to extra symmetries which are not isometries of the squashed manifold.

Given that the kappa-symmetry gauge choice allows only for certain string configurations, it would be interesting to study integrability in a more general setting. It would also be enlightening to look at the worldsheet theory using the near \textsc{bmn} expansion as it was done for the $\AdS_{3}\times S^{3} \times S^{3}\times S^{1}$ background in~\cite{Rughoonauth:2012qd}.

\subsection*{Acknowledgements}
We are indebted to A. Tseytlin and S. Reffert for discussion and an attentive revision of the manuscript.
L.I.U. would like to thank the Theory Division of CERN for hospitality during the final stages of this work. L.I.U. is supported by a STFC Postdoctoral Research Fellowship.

\appendix
\section{Conventions}
\label{appA:title}
\subsection{Gamma Matrices}
\label{appA:gamma}
The bosonic subalgebra of $psu(1,1\vert 2)$ consists of
two commuting $\mathfrak{sl}(2)$, with one of the $sl(2)$
non-compact and the other compact, so that the bosonic
subalgebra is $\mathfrak{sl}(2,\mathbbm{R})\oplus\mathfrak{su}(2)$.

To describe the action of the $\mathfrak{sl}(2)$ generators on the
supercharges we introduce the following sets of gamma matrices:
\begin{align}\label{3ddirac}
  \gamma^\mu &=(i\sigma ^2,\sigma ^1,\sigma ^3) \, , & 
  \gamma^n &= (\sigma ^1,\sigma ^2,\sigma ^3) \,,& 
  \gamma^{\dot{n}} &= (\sigma^1,\sigma ^2,\sigma ^3).
\end{align}
We can choose the following representation for the 10d Dirac matrices:
\begin{align}\label{gamma1}
  \Gamma^\mu &= \sigma^1\otimes \sigma ^2\otimes \gamma ^\mu\otimes
  \mathbbm{1} \otimes \mathbbm{1}\,, &\mu=0,1,2 \\
  \label{gamma2} 
  \Gamma^{n} &= \sigma^1\otimes
\sigma^1\otimes \mathbbm{1} \otimes \gamma ^{n} \otimes \mathbbm{1}\,, & n=3,4,5\\
\label{gamma3}
\Gamma^{\dot{n}} &= \sigma^1\otimes \sigma^3 \otimes \mathbbm{1} \otimes \mathbbm{1}
\otimes \gamma ^{\dot{n}}\,,& \dot{n}=6,7,8\\
\Gamma^9 &= -\sigma^2\otimes \mathbbm{1} \otimes \mathbbm{1} \otimes
\mathbbm{1} \otimes \mathbbm{1},
\end{align}
where the 3d gamma-matrices $\gamma ^i$ are taken from
(\ref{3ddirac}).
 The charge conjugation matrix is given by
\begin{equation}\label{chargeconjmatr}
 C=i\sigma ^2\otimes\sigma ^2\otimes\sigma ^2\otimes\sigma ^2\otimes\sigma
 ^2.
\end{equation}

\subsection{Vielbein}
From \eqref{eq:IIB-metric}, we can extract the vielbein. Before changing coordinates, we have
\begin{align}
\begin{array}{lll}
e^{0}=R\cosh \omega \di \tau &\quad e^{1}=R \di \omega &\quad e^{2}=R\sinh{\omega}\di \tau+R\di \sigma \\
e^{3}=R\di \theta &\quad e^{4}=R\sin(\theta)\di \phi &\quad e^{5}=R\cos{\theta}\di \phi
+Rd\psi \\
e^{i}= 2R \cot \varpi \di y_i &\quad(i=6,\cdots, 9) \ .
\end{array}
\label{originalvielbein}
\end{align}
After we change coordinates to $\psi=\alpha+2y_{9}$ the vielbein becomes:
\begin{align}
\begin{array}{lll}
e^{3}=R\di \theta &\quad e^{4}=R\sin{\theta}\di\phi &\quad e^{5}=R \cos{\theta}\di\phi
+R\di \alpha  +2R \di y_{9}\ , \\
\end{array}
\end{align}
and the T--dual vielbein ${\tilde{e}_{M}}^{a}$ reads
\begin{align}
  \wt{e}^{0} &= R \cosh(\omega)\di \tau \qquad   \wt{e}^{1} = R \di \omega \qquad  \wt{e}^{2}  = R \sinh{\omega}\di \tau+R\di \sigma \nonumber \\
  \wt{e}^{3} &= R \di \theta \qquad \qquad \wt{e}^{4} = R
  \sin(\theta)\di \phi \qquad  \wt{e}^{i} = 2R \tan \varpi \di y_i 
  \qquad (i=6,\cdots, 8) \nonumber \\
  \wt{e}^{5} &=
  \frac{\cos^2 \varpi}{2R} \di \wt y_9 + R \sin^2{\varpi} (\di
  \alpha+\cos{\theta}\di \phi) \nonumber \\ 
   \wt{e}^{9} &=
   \cos\varpi\sin\varpi\left( \frac{\di \wt
       y_9}{2R} - R \left( \di \alpha+\cos{\theta}\di \phi
     \right)\right) \ .
\label{tdualvielbein}
\end{align}
We can write down the gamma matrices in the coordinate frame $\Gamma_{M}$ in terms of the gamma matrices $\gamma_{a}$ in the orthogonal frame (tangent space). 
\begin{align}
  \Gamma_{\tau} &= R \cosh{\omega}\gamma_{0}+R\sinh{\omega}\gamma_{2} & \Gamma_{\omega} &= R\gamma_{1}  \nonumber \\
   \Gamma_{\sigma} &= R\gamma_{2}  &
\Gamma_{\theta} &= R\gamma_{3} \nonumber \\
\Gamma_{\phi} &= R\sin{\theta}\gamma_{4}+R\cos{\theta}\sin^2 \varpi \gamma_{5}-R\cos{\theta}\cos\varpi\sin\varpi\gamma_{9}
   &   \Gamma_{\zeta} &= \cos \varpi \gamma_{5}+\sin \varpi \gamma_{9} \nonumber \\ 
  \Gamma_{\alpha} &= R\sin^{2}\varpi\gamma_{5}-R\sin\varpi\cos\varpi\gamma_{9} &
   \Gamma_{i} &= 2R\tan \varpi \gamma_{i} 
\end{align}
The gamma matrices $\Gamma^M $ read:
\begin{align}
\Gamma^{\tau} &= \frac{\mathrm{sech}{\omega}}{R}\gamma^{0} & \Gamma^{\omega} &= \frac{1}{R} \gamma^{1}  & \Gamma^{\sigma} &= \frac{1}{R}(-\tanh{\omega}\gamma^{0}+\gamma^{2}) \nonumber \\
 \Gamma^{\theta} &= \frac{1}{R}\gamma^{3} & \Gamma^{\phi} &= \frac{1}{R}\csc{\theta}\gamma^{4} &\Gamma^{i}
    &= \frac{\cot\varpi}{2R} \gamma^{i} \nonumber \\
    \Gamma^{\alpha} &= \frac{1}{R}(\gamma^{5}-\cot{\theta}\gamma^{4}-\cot \varpi \gamma^{9}) &
    \Gamma^{\zeta} &= \cos\varpi\gamma^{5}+\sin\varpi\gamma^{9} \ ,
\end{align}
and the spin connection is
\begin{equation}
  \begin{aligned}
    \slashed{\omega}_{\tau}& = -\sinh{\omega}\gamma^{01}+\cosh{\omega}\gamma^{12}  \\
    \slashed{\omega}_{\sigma}&=-\gamma^{01}   \\
    \slashed{\omega}_{\omega}&=-\gamma^{02}  \\
    \slashed{\omega}_{\theta}&=-\sin^2 \varpi \gamma^{45}+\sin\varpi\cos\varpi\gamma^{49}  \\
    \slashed{\omega}_{\phi}&=-\cos{\theta} \left( 1 +
      \cos^2\varpi \right)\gamma^{34}+\sin^2\varpi \sin{\theta}\gamma^{35}-\sin\varpi\cos\varpi\sin{\theta}\gamma^{39}  \\
    \slashed{\omega}_{\alpha}&=\sin^2 \varpi\gamma^{34}  \\
    \slashed{\omega}_{i}&=0 & (i=6, \cdots ,9)
  \end{aligned}
 \label{spinconn}
\end{equation}
Using $\zeta$, the vielbein components $\wt{e}^{(5)}$ and $\wt{e}^{(9)}$ in \eqref{tdualvielbein} can be rewritten as:
\begin{equation}
  \begin{aligned}
    \wt{e}^{5}&=\cos\varpi \di \zeta+R\sin^2 \varpi(\di\alpha+\cos\theta\di\phi) \ , \\
    \wt{e}^{9}&=\sin\varpi\di\zeta-R\cos\varpi\sin\varpi(\di\alpha+\cos\theta\di\phi) \ .
  \end{aligned}
\label{newvielbcomp}
\end{equation}

\printbibliography

\end{document}